\documentstyle[onecolumn,epsf,mnras_cite]{mn}
\input{epsf}
\newcommand{\ba}{\begin{eqnarray}}
\newcommand{\ea}{\end{eqnarray}}
\newcommand{\be}{\begin{equation}}
\newcommand{\ee}{\end{equation}}
\newcommand{\br}{{\bf r}}
\newcommand{\bs}{{\bf s}}
\newcommand{\bk}{{\bf k}}
\newcommand{\bq}{{\bf q}}
\newcommand{\eps}{\epsilon}
\newcommand{\nn}{\nonumber \\}
\newcommand{\ls}{\mathrel{\raise1.16pt\hbox{$<$}\kern-7.0pt 
\lower3.06pt\hbox{{$\scriptstyle \sim$}}}}         
\newcommand{\gs}{\mathrel{\raise1.16pt\hbox{$>$}\kern-7.0pt 
\lower3.06pt\hbox{{$\scriptstyle \sim$}}}}         

\title{The nonlinear redshift-space power spectrum of galaxies}
\author[A.F. Heavens, S. Matarrese, L. Verde]
{A.F. Heavens$^{1}$, S. Matarrese$^{2}$, L. Verde$^{1}$\\
$^{1}$ Institute for Astronomy, University of Edinburgh, Royal Observatory, 
Blackford Hill, Edinburgh EH9 3HJ , United Kingdom\\
$^{2}$ Dipartimento di Fisica {\em Galileo Galilei}, Universit\`{a} di
Padova, via Marzolo 8, I-35131 Padova, Italy\\}

\begin{document}
\maketitle
\begin{abstract}
We study the power spectrum of galaxies in redshift space, with 
third order perturbation theory to include corrections that are 
absent in linear theory. 
We assume a local bias for the galaxies: i.e.
the galaxy density is sampled from some local function of the 
underlying mass distribution.  
We find that the effect of the nonlinear bias in real space is to 
introduce two new features:  first, there is a contribution to the power 
which is constant with wavenumber, whose nature we reveal as 
essentially a shot-noise term.  In principle this contribution can mask
the primordial power spectrum, and could limit the accuracy with which the
latter might be measured on very large scales.  Secondly, the effect of 
second- and third-order bias is to modify the effective bias (defined as 
the square root of the ratio of galaxy power spectrum to matter power 
spectrum).  The effective bias is almost scale-independent over a wide range 
of scales.  
These general conclusions also hold in redshift space.  In addition, we 
have investigated the distortion of the power spectrum by peculiar velocities,
which may be used to constrain the density of the Universe. We look at the
quadrupole-to-monopole ratio, and find that higher-order terms can mimic 
linear theory bias, but the bias implied is neither the linear bias, nor
the effective bias referred to above.
We test the theory with biased N-body simulations, and find excellent 
agreement in both real and redshift space, 
providing the local biasing is applied on a scale whose fractional r.m.s.
density fluctuations are $< 0.5$.
\end{abstract}
 


\section{Introduction}

The clustering properties of galaxies are dependent to a certain extent on the 
precise population analysed.  This implies that not all populations can be
unbiased tracers of the matter distribution - at least one must be biased.
Simple bias models have been in existence since the early Cold Dark Matter
(CDM) simulations suggested that pairwise velocities of galaxies would be too
high unless the (optically-selected) galaxies were more strongly clustered 
than the mass \cite{DEFW85}.  In the first simple models, it was assumed that
the fractional overdensity in galaxies $\delta^g \equiv 
\delta n/n$ was simply a multiple of the mass overdensity:
$\delta^g = b\delta$.  This linear bias model cannot be true in detail for
all populations, since the shape of the power spectrum is unchanged in this
case, and not all galaxy populations have the same shape of power spectrum,
although the differences are not large \cite{PD94}.  At a more fundamental 
level, such a model could only survive if applied to a smoothed field,
otherwise $\delta^g < - 1$ which corresponds to a negative galaxy density.
The linear bias model can be viewed as an approximation to a more general
Eulerian bias prescription, where the galaxy density is assumed to be some
function of the present-day mass density (e.g. \pcite{Coles93};
\pcite{CCMM94}; \pcite{Wein95}; \pcite{MPH98}).  The
linear bias term is then the first interesting term in a Taylor expansion 
of the function about $\delta=0$.   If the galaxy distribution is significantly
biased with respect to the matter, there are major implications
for cosmology, arising from the difficulty in detecting the matter 
distribution.   On scales where masses can be reliably measured, such as in 
galaxy clusters, bias is required to reconcile the mass-to-light ratios
with an Einstein-de Sitter Universe.  Separately, linear studies of peculiar 
velocities, either directly (e.g. \pcite{SEDSY98}) or via redshift distortions
(e.g. \pcite{Ham92}, \pcite{HT95}; \pcite{CFW95}; \pcite{HBCJ95}, 
\pcite{Ham98}, \pcite{HC98}) return the density 
parameter of the Universe only via $\beta \equiv \Omega_0^{0.6}/b$, 
so ignorance of $b$ compromises determinations of $\Omega_0$.
In addition to the Eulerian bias discussed in this paper, there are
other possibilities, such as Lagrangian bias (e.g. \pcite{DEFW85}, 
\pcite{BBKS}) and stochastic bias (\pcite{Pen97}; \pcite{TegPeeb98}, 
\pcite{DL98}), biasing determined by
halo properties at some redshift (\pcite{MW96}, \pcite{CLMP98}, 
\pcite{CMP98}), or 
biasing determined by coherent processes over a large scale (\pcite{BW91};
\pcite{BCFW93}).  
Despite a major industry in modelling (e.g. \pcite{KCDW98}), it is 
probably fair to say that we still have a great deal of 
uncertainty in precisely
where galaxies should form, so it is not clear which, if any, of the
above descriptions corresponds closely to reality.  In this paper, we 
assume an Eulerian description for the bias, but we do not restrict it to
linear form.  

The current generation of galaxy redshift surveys, such as the Anglo-Australian
Telescope 2dF survey \cite{Colless96}, and the Sloan Digital Sky Survey 
\cite{GW95} have sufficient space density and volume to look beyond linear
theory, and this opens up possibilities of lifting the degeneracy between
$\Omega_0$ and $b$, by measuring $b$ empirically through higher-order 
statistics (\pcite{MVH97}; \pcite{VHMM98}).   
Higher-order studies of the power 
spectrum have been made in real space by \scite{MSS92}, \scite{JB94} and
\scite{BE94b}, and 
also in an elegant treatment in real and redshift space in the Zel'dovich 
approximation by \scite{TayHam96}.   Our treatment differs from previous 
perturbative calculations by its inclusion of bias and redshift distortions,
and from the Zel'dovich approximation by inclusion of nonlinear, local bias 
and by the different treatment of nonlinear evolution.  The layout of 
the paper is as follows:  in Section 2 we set out the calculations formally,
treating carefully the transition from real to redshift space;
in Sections 3 and 4 we separate the major effects which come into operation 
beyond linear theory, in real and redshift space, and in Section 5 we 
present our conclusions.

\section{Method}

\subsection{Real to redshift space}

Let \br\ and \bs\ be the real and redshift-space coordinates, with
the observer at the origin.  The latter is defined to be the recession velocity
(including the peculiar velocities {\bf v} of galaxy and observer) 
from the observer,
divided by the Hubble constant $H_0$.  $H_0=1$ is equivalent to using 
recession velocity as the distance coordinate, and we assume this from now on.
Further we define $\rho_r(\br)$ and $\rho_s(\bs)$ to be the density fields
in real and redshift space.   The mean density may be spatially-dependent,
because of selection effects;  we define the expected densities as 
$\phi_r(\br)$ and $\phi_s(\bs)$.  The overdensity in redshift space is
defined by $1+\delta_s(\bs) \equiv \rho_s(\bs)/\phi_s(\bs)$, and similarly for
the real-space overdensity $\delta_r$.  In all, we shall work with four 
random fields:
\begin{itemize}
\item{$\delta_r$:  mass overdensity in real space}
\item{$\delta_s$:  mass overdensity in redshift space}
\item{$\delta^g_r$:  galaxy overdensity in real space}
\item{$\delta^g_s$:  galaxy overdensity in real space}
\end{itemize}
and each will have its corresponding power spectrum $P^g_s$ etc.  For 
simplicity and consistency with previous notation, we drop the $r$ subscript 
on the real space mass power spectrum.

The coordinate transform from real to redshift space is \cite{Kai87}
\be
\bs(\br) = \br\left[1+{U(\br)-U({\bf 0})\over r}\right]
\ee
where $U(\br) \equiv {\bf v}\cdot \br /(H_0 r)$.  Number conservation implies
$\rho_r(\br)d^3\br = \rho_s(\bs)d^3\bs$, which gives
\be
\left[1+\delta_s(\bs)\right]\phi_s(\bs) = 
{\left[1+\delta_r(\br)\right]\phi_r(\br)\over J} 
\ee
where the Jacobian is
\be
J = \left[1+{\Delta U\over r}\right]^2 \left(1+{\partial U(\br)
\over \partial r}\right).
\ee
and $\Delta U(\br)\equiv U(\br)-U({\bf 0})$.  
We make now the large-distant-volume approximation, where we assume that any 
modes we analyse have wavenumbers $k$ which satisfy $kr \gg 1$ throughout. 
Terms $\Delta U/r$, which are $\sim \delta/(kr)$ if $\Omega \simeq 1$,
are ignored entirely in comparison with $\delta$.   If we assume that
$\phi$ drops as some power of $r$, then a Taylor expansion of $\phi(s)$ yields
$\phi(r)$ plus negligible correction terms.
Beyond linear theory,  the lowest-order contributions to the power spectrum
arise from both second- and third-order terms in $\delta$,  so we expand to
third order:
\be
\delta_s(\bs) = \delta_r(\br) - U'(\br)+U'^2(\br) - \delta_r(\br) U'(\br) + 
\delta_r(\br)U'^2(\br) - U'^3(\br) \equiv F(\br),
\label{deltas}
\ee
and $' \equiv \partial/\partial r$.
To linear order,  differences in the argument do not matter, but to third 
order, they do.  To the required order,
\be
F(\br) = F(\bs) - \Delta U(\bs) F'(\bs) \left[1-U'(\bs)\right] +
{1\over 2} \Delta U^2(\bs) F''(\bs),
\ee
which gives us the final expression for the redshift-space overdensity in terms
of the real-space overdensity, all evaluated at \bs:
\be
\delta_s  =  \delta_r - U'+U'^2 - \delta_r U' + 
\delta_r U'^2 - U'^3 - \Delta U \delta_r' + \Delta U U'' - 
3 \Delta U U' U'' + 2 \Delta U U' \delta_r' + 
\Delta U U'' \delta_r + {1\over 2} \Delta U^2 \delta_r'' - {1\over 2} 
\Delta U^2 U'''.
\ee

\subsection{Bias and evolution}

We assume that the galaxy overdensity field $\delta^g \equiv \delta n/n$, 
where $n$ is the number density of galaxies, is related to the mass
overdensity field  $\delta$ via a local
function.  This plausible `local Eulerian' bias model is chosen for simplicity
and tractability, and has some support from simulation (e.g. \pcite{CO92}, 
\pcite{KNS97}).  
Of course other schemes are possible, such as Lagrangian 
bias (e.g. \pcite{CLMP98}; see also \pcite{CK89}, \pcite{MW96}), or nonlocal
bias (e.g. \pcite{BW92}, \pcite{BCFW93}, \pcite{Matsubara95}).  
We expand the local function as a Taylor series around $\delta_r=0$ 
\cite{FG93}:
\be
\delta^g({\bf r}) = \sum_{j=0}^{\infty} {b_j\over j!} \,\delta_r^j.
\label{deltag}
\ee
An unbiased galaxy field would have $b_1=1$ and all other bias parameters zero.
Already there is a subtlety.  We will truncate the expansion,
which will only be a good approximation if the value of $\delta$ is 
typically much less than unity.  Therefore our biasing assumption is that 
there is some smoothing scale for which (\ref{deltag}) is a good approximation,
and the fields above should be interpreted as the smoothed galaxy and mass
density fields.   For the perturbative expansion to be valid, we choose a 
smoothing scale large enough that the smoothed density field has small 
fluctuations (this will be quantified later).  If we wish to smooth on a 
smaller scale, then a numerical approach is probably necessary \cite{MPH98}.
Note also that the galaxy distribution is of course a point
process.  We assume that the positions are determined by a Poisson sampling 
of a density field whose overdensity is $\delta_g$.   We do not consider 
stochastic biasing \cite{Pen97}, and ignore shot noise in the power spectrum.

The next non-zero terms beyond linear theory will be of order 4 in $\delta_r$,
so we need to keep terms up to $j=3$.  Throughout we ignore the $b_0$ term, 
required to ensure that $\langle \delta^g_r\rangle=0$, since we
will be interested in the spectral properties of the galaxy field, and
drop all terms which contribute only to ${\bf k}={\bf 0}$. 

Inserting (\ref{deltag}) into (\ref{deltas}), we get, to third order
in $\delta_r$:
\ba
\delta^g_s & = & \left[b_1\delta_r - U'\right]+\left[{b_2\over 2} 
\delta_r^2+U'^2 -b_1\delta_r\,U' - b_1\delta_r'\Delta U 
+ \Delta U\,U''\right]
+ \left[{b_3\over 6}\delta_r^3 - {b_2\over 2}\delta_r^2 
U' + b_1\delta_r\, U'^2 
- U'^3 - b_2 \Delta U \delta_r\delta_r' \right.\nn
& & \left.-3 \Delta U\,U'U''+2b_1\delta_r'\Delta U\,U' + 
b_1\delta_r \Delta U\,U'' + {1\over 2}b_1 \delta_r''(\Delta U)^2 
- {1\over 2}(\Delta U)^2 U'''\right].\nn
\label{deltas3}
\ea
The Fourier transform of this may be taken in redshift space, 
making use of the transforms of simple products:
\ba
(XY)_\bk & = & 
{1\over (2\pi)^3}\int d^3\bk_1 d^3\bk_2 \delta^D(\bk-\bk_1-\bk_2)
X_{1} Y_{2}\nn
(XYZ)_\bk & = & 
{1\over (2\pi)^6}\int d^3\bk_1 d^3\bk_2 d^3\bk_3 \delta^D(\bk-\bk_1-\bk_2
-\bk_3)
X_{1} Y_{2} Z_3\nn
\ea
where $X_1$ is the $\bk_1$ component of $X_\bk$ etc, and $\delta^D$ is the
Dirac delta function.   In the distant-observer
approximation, $\partial/\partial r \rightarrow ik\mu$, where $\mu \equiv
\bk\cdot \hat \br/k$ and $\hat \br$ is a unit vector from the observer to the
galaxy (assumed constant across the sample)
and the transform
of $U$ is $U_\bk = i\mu f\eta_\bk/k$, where $-\eta_\bk$ is 
the transform of the velocity divergence. $f \equiv d\ln D/d\ln a
\simeq \Omega_0^{0.6}$, where $D(a)$ is the growing-mode amplitude and $a$ 
is the scale factor of the Universe.  Second-order terms were computed by
\scite{Peebles};
the third-order expansion of the fluid equations (see \pcite{Fry84}) 
is detailed in
\scite{CM94a} and \scite{CM94b}, giving $\delta_\bk$:
\ba
\delta_\bk & = &  \eps_\bk + 
{1\over (2\pi)^3}\int d^3\bk_1 d^3\bk_2 \delta^D(\bk-\bk_1-\bk_2)
J^{(2)}_S(\bk_1,\bk_2)
\eps_1 \eps_2 \nn
& + &{1\over (2\pi)^6}\int d^3\bk_1 d^3\bk_2 d^3\bk_3 
\delta^D(\bk-\bk_1-\bk_2-\bk_3)J^{(3)}_S(\bk_1,\bk_2,\bk_3)\eps_1 \eps_2\eps_3.
\nn
\ea
$\eps_\bk$ is the linear, real-space Fourier coefficient of the
density field $\delta(\br)$.  The corresponding expressions for 
$\eta_\bk$ involve 
replacing $J$ by $K$.  The functions $J$ and $K$ are quoted for 
$\Omega_0=1$,  as computed by \scite{Goroff86},  \scite{CM94a} and 
\scite{CM94b}:
\ba
J^{(2)}_S(\bk_1,\bk_2)&  = & {5\over 7} + {\bk_1\cdot \bk_2\over 2 k_1 k_2}
\left({k_1\over k_2} + {k_2\over k_1}\right) + {2\over 7}\left({\bk_1\cdot
\bk_2\over k_1 k_2}\right)^2,\nn
K^{(2)}_S(\bk_1,\bk_2)&  = & {3\over 7} + {\bk_1\cdot \bk_2\over 2 k_1 k_2}
\left({k_1\over k_2} + {k_2\over k_1}\right) + {4\over 7}\left({\bk_1\cdot
\bk_2\over k_1 k_2}\right)^2,\nn
J^{(3)}(\bk_1,\bk_2,\bk_3)& = & J^{(2)}(\bk_2,\bk_3)
\left[{1\over 3} + {1\over 3}{\bk_1\cdot(\bk_2+\bk_3)\over (\bk_2+\bk_3)^2}
+{4\over 9}{\bk\cdot\bk_1\over k_1^2}{\bk\cdot(\bk_2+\bk_3)\over
(\bk_2+\bk_3)^2}\right]\nn
&&-{2\over9} {\bk\cdot\bk_1\over k_1^2}{\bk\cdot(\bk_2+\bk_3)\over
(\bk_2+\bk_3)^2}{\bk_3\cdot(\bk_2+\bk_3)\over
k_3^2} + {1\over 9} {\bk\cdot\bk_2\over k_2^2} {\bk\cdot\bk_3\over k_3^2}\nn
K^{(3)}(\bk_1,\bk_2,\bk_3)& = & 3J^{(3)}(\bk_1,\bk_2,\bk_3) - {\bk\cdot
\bk_1\over k_1^2}J^{(2)}(\bk_2,\bk_3)- {\bk\cdot(\bk_1+\bk_2)\over 
(\bk_1+\bk_2)^2}K^{(2)}(\bk_1,\bk_2),\nn
\ea
with $\bk=\bk_1+\bk_2+\bk_3$ in the last two expressions.  The subscript $S$ 
indicates that the expression has been made symmetric w.r.t. any permutation
of the arguments.  If not, then the symmetrized kernel must be obtained by
averaging the quoted expression over all permutations.
These kernels are correct for $\Omega_0=1$ but are only 
weakly dependent on $\Omega_0$ 
(e.g. \pcite{BJCP92}, \pcite{CLMM95}; see also \pcite{Bernardeau94b}, 
\pcite{Eis97}, \pcite{KB98}).
After some manipulations,
the transform of the third-order, biased, redshift-space density field is
\ba
\delta^g_{s\bk} & = & F^{(1)}_S(\bk) \eps_\bk +{1\over (2\pi)^3}\int d^3\bk_1 
d^3\bk_2 \delta^D(\bk-\bk_1-\bk_2)
F^{(2)}_S(\bk_1,\bk_2)\eps_1\eps_2\nn
& & +{1\over (2\pi)^6}\int d^3\bk_1 d^3\bk_2 d^3\bk_3 
\delta^D(\bk-\bk_1-\bk_2-\bk_3)
F^{(3)}_S(\bk_1,\bk_2,\bk_3)\eps_1\eps_2\eps_3\nn
\ea
with the following kernels:
\ba
F^{(1)}_S(\bk) & = & b_1 + f\mu^2\nn
F^{(2)}_S(\bk_1,\bk_2) & = & b_1 J_S^{(2)}(\bk_1,\bk_2) + f\mu^2 
K^{(2)}_S(\bk_1,\bk_2) + {1\over 2}b_2 +  {b_1 f\over 2} 
\left[\mu_1^2+\mu_2^2 + \mu_1\mu_2\left(
{k_1\over k_2}+{k_2\over k_1}\right)\right]\nn
& & + f^2\left[\mu_1^2\mu_2^2+{\mu_1\mu_2\over 2}\left(
\mu_1^2{k_1\over k_2}+\mu_2^2{k_2\over k_1}\right)\right]\nn
F^{(3)}(\bk_1,\bk_2,\bk_3) & = & b_1 J^{(3)}(\bk_1,\bk_2,\bk_3)
+ f\mu^2 K^{(3)}(\bk_1,\bk_2,\bk_3)
+ {b_2\over 2}f\mu_3^2 
+ {b_3\over 6}
+ {b_2\over 2}f\mu_1\mu_2{k_2\over k_1}
+ {b_2\over 2}f\mu_1\mu_3{k_3\over k_1}
+ b_1 f^2 \mu_2^2\mu_3^2  \nn
& + & 2b_1 f^2 \mu_1 \mu_2 \mu_3^2{k_1\over k_2}
+ b_1 f^2 \mu_2 \mu_3^3 {k_3\over k_2}
+ {b_1\over 2}f^2 \mu_1^2\mu_2\mu_3 {k_1^2\over k_2 k_3}
+ f^3 \mu_1^2\mu_2^2\mu_3^2
+ 3 f^3  \mu_1\mu_2^2\mu_3^3 {k_3\over k_1}
+ {1\over 2}f^3 \mu_1\mu_2\mu_3^4 {k_3^2\over k_1 k_2}\nn
& + &  J^{(2)}(\bk_2,\bk_3)\left(b_2 + b_1 f \mu_1^2 + b_1 f 
\mu_1\mu_{2+3}{k_{2+3}\over k_1}\right)\nn
& + & K^{(2)}(\bk_2,\bk_3)\left(b_1 f  \mu_{2+3}^2
     + b_1 f \mu_1 \mu_{2+3} {k_1\over k_{2+3}} 
     + 2 f^2 \mu_1^2\mu_{2+3}^2
     + f^2 \mu_1\mu_{2+3}^3 {k_{2+3}\over k_1}
     + f^2  \mu_1^3\mu_{2+3} {k_1\over k_{2+3}}\right).\nn
\ea 
A subscript $S$ on
the $F$ terms also indicates that the term has been symmetrized w.r.t. its
arguments. Note that the last term has not been symmetrized and $\mu_{2+3}
\equiv (\bk_2+\bk_3).\hat \br/|\bk_2+\bk_3|$.

From these equations we can obtain the power spectrum to third-order ,
which includes non-zero correction terms of two types, as in the real-space
unbiased case (\pcite{MSS92}; \pcite{JB94}).  The redshift-space power 
spectrum $P^g_s(\bk)$ is defined by
\be
\langle \delta^g_{s\bk_1} \delta^g_{s\bk_2}\rangle = (2\pi)^3 P^g_s(\bk_1)
\delta^D(\bk_1+\bk_2)
\ee
and the real-space mass linear power spectrum $P_{11}(k)$ is defined 
similarly by 
$\langle \eps_{\bk_1} 
\eps_{\bk_2}\rangle = (2\pi)^3 P_{11}(k_1) \delta^D(\bk_1+\bk_2)$.  The 
Gaussian nature of the initial fluctuations implies, by Wick's theorem, that
$\langle \eps_1 \eps_2 \eps_3 \rangle = 0$ and 
$\langle \eps_1 \eps_2 \eps_3 \eps_4 \rangle = 
\langle \eps_1 \eps_2 \rangle \langle \eps_3 \eps_4 \rangle$ plus cyclic
permutations.   Hence we find two (one-loop) terms which arise at the next 
level beyond linear theory for the spectrum (tree level):
\ba
P^g_s(\bk) & \equiv & P^g_{s11} + P^g_{s22}+P^g_{s13}\nn
&  = & (1+\beta\mu^2)^2 b_1^2 P_{11}(k) +  
2\int\,{d^3\bq\over (2\pi)^3}P_{11}(q)P_{11}(|\bk-\bq|)\left[F_S^{(2)}
(\bq,\bk-\bq)\right]^2\nn  
& + & 6 (1+\beta\mu^2) b_1 P_{11}(k) \int\,{d^3\bq\over (2\pi)^3}P_{11}(q)
F_S^{(3)}(\bq,-\bq,\bk),\nn
\ea
where $\beta\equiv f/b_1$. In the unbiased case in real space ($\beta=0$), 
these formulae reduce to 
those in \scite{JB94} (note there is a factor of $k^2$ missing in their 
equation 19).  For this case, the limits of integration
for the $P^g_{22}$ and $P^g_{13}$ terms are not too important, provided the
power spectrum decays sufficiently rapidly at high $k$ (faster than $k^{-1}$),
and this is fine for spectra of practical interest.    For the bias terms,
however, one has to be careful, because if one places a cutoff in the 
integrals, one finds that the resulting power spectrum is sensitive to
this cutoff.   This should not be surprising:  the cutoff effectively 
smoothes the linear power spectrum (with a top-hat filter in $k-$space), and
the power spectrum depends on what scale the filtering takes place.  
Furthermore, it makes physical sense:  if a local bias prescription 
applies at all, it will be necessary to specify on what scale it acts.
These considerations lead us to the following Eulerian bias proposal:
\begin{itemize}
\item{Evolve $\delta$ to 3rd-order}
\item{Smooth with some filter function corresponding to a smoothing scale
$R_f$}
\item{Apply the Taylor expansion to third-order to get $\delta_g$.}
\end{itemize}
The resulting field is evidently itself smoothed.  This procedure is 
effected by replacing all occurrences of the linear $\eps_k$ by 
$\eps_k W_k$, where $W_k$ is the transform of the filter function
[e.g. $W_k = \exp(-k^2R_f^2/2)$ for a Gaussian filter].  We shall 
not explicitly retain these factors of $W_k$; the smoothing is
implicit in $\eps_k$.  Note that the operations of smoothing and biasing
do not commute;   unless the matter power spectrum has a natural 
small-scale cutoff, the order assumed here is the only sensible choice
for a perturbative calculation.  This is a critical assumption, of course,
but we reiterate that it has some support from simulation (e.g. \pcite{CO92}). 
For further discussions of the effects
of smoothing, see \scite{CMS93}, \scite{MS93}, \scite{Bernardeau94a}.

\subsection{Bispectrum}

The galaxy redshift-space bispectrum $B^g_s(\bk_1,\bk_2,\bk_3)$ is the 
3-point function in Fourier space, defined by
\be
\langle \delta_{s\bk_1} \delta_{s\bk_2} \delta_{s\bk_3}\rangle = 
(2\pi)^3 B^g_s(\bk_1,\bk_2,\bk_3)
\delta^D(\bk_1+\bk_2+\bk_3).
\ee
In real-space, this has been studied by \scite{MVH97}, \scite{Sco97} and
\scite{SCFFHM98}.  
We can use the preceding formalism to obtain the biased, redshift-space
bispectrum, in the distant observer approximation:
\ba
B^g_s(\bk_1,\bk_2,\bk_3) & = & 2 P_{11}(k_1)P_{11}(k_2)F_S^{(1)}(\bk_1)
F_S^{(1)}
(\bk_2)F_S^{(2)}(\bk_1,\bk_2){\rm \ +\ cyclic\ terms\ (1,2)\rightarrow
(2,3)\ and\ (3,1)}\nn
& = & 2 P_{11}(k_1) P_{11}(k_2)b_1^3(1+\beta\mu_1^2)(1+\beta\mu_2^2)
\left[J^{(2)}(\bk_1,\bk_2) + 
\beta\mu_2 K^{(2)}(\bk_1,\bk_2) + {b_2\over 2b_1}
+ b_1\beta^2\mu_1^2\mu_2^2 + {b_1\beta\over 2}(\mu_1^2+\mu_2^2)\right.\nn
& + & \left.{b_1\beta\over 2}\mu_1\mu_2\left({k_1\over k_2}+{k_2\over k_1}
\right)
+ {b_1^2\over 2}\beta^2\mu_1\mu_2\left(\mu_1^2{k_1\over k_2}+
\mu_2^2{k_2\over k_1}\right)\right]{\rm \ +\ cyclic\ terms}.\nn
\ea
In a separate paper \cite{VHMM98} we use this to investigate how redshift
surveys can be used practically to estimate $b_1$ and hence determine
the density parameter (via $\beta$).

\section{New features: real space}

Interestingly, there are some new features which arise in real-space 
from biasing alone.  Consider the real-space corrections to the
power spectrum, $P^g_{22} \equiv P^g_{s22}(\beta=0)$ and $P^g_{13} \equiv 
P^g_{s13}(\beta=0)$.   
At small $k$, the former is more-or-less constant with k,
while the latter is proportional to $P_{11}(k)$.  
We look at each of these in turn.

\begin{figure}
\centering
\begin{picture}(250,250)
\includegraphics{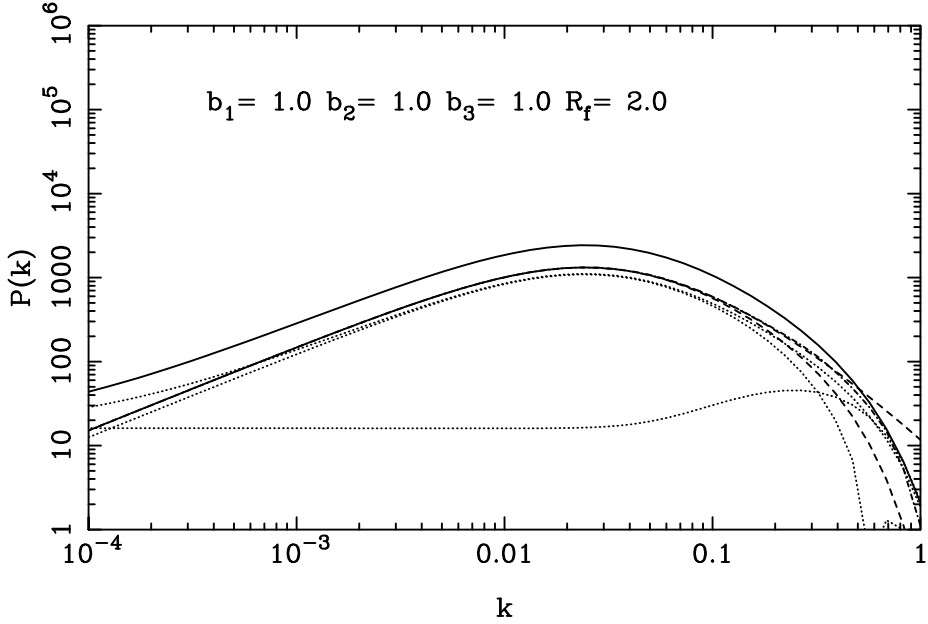}
\end{picture}
\caption[]{\label{RealPower} Real-space power spectrum.  The (upper) 
solid line is the
final power spectrum;  the dashed lines are the linear power spectrum,
unsmoothed and smoothed, and the dotted lines are (from the bottom at 
$k=10^{-3}$) the contributions to the
power from $P_{22}$, $P_{13}$ and their sum.  The dot-dashed line is the 
second-order mass power spectrum, which merges with the dashed linear 
power spectrum to form the lower solid line.  For model details, see text.}
\end{figure}
\subsection{Constant power on very large scales}

On very large scales, where $\Delta^2(k)\ll1$ (where 
$\Delta^2$ is the contribution to the variance in mass overdensity per unit 
$\ln(k)$), the terms arising from 
nonlinear evolution (and a linear bias $b_1$) are generally small
\cite{JB94}, so we concentrate on the contributions from nonlinear
bias.   

The $P^g_{22}$ term is
\ba
P^g_{22}(k) & = & {1\over 4\pi^3} \int d^3 \bk_1 P(\bk_1)P(|\bk-\bk_1|)\left[
b_1 J^{(2)}_s(\bk_1,\bk-\bk_1)+{b_2\over 2}\right]^2\nn
\label{DC}
\ea
which provides a constant contribution to the power when $|\bk|$ is small.  
The presence of such
a term was suggested by \scite{Coles93} and \scite{SW95};  
here we provide a mechanism for calculating it perturbatively 
for arbitrary local bias.

For an underlying matter power spectrum which falls to zero as 
$k\rightarrow 0$, $P^g_{22}$ will eventually dominate over both the linear 
term and $P^g_{13}$, leading to a divergent
effective bias $b^2_{\rm eff}(k) \equiv P^g(k)/P(k)$ in this limit ($P$
is the matter power spectrum evolved to third order;  on these large scales,
linear theory is accurate, so $P \simeq P_{11}$).
In turns out, for an underlying Zel'dovich spectrum, this term is unlikely 
to be of importance unless the other correction term $P^g_{13}$ below is
small.    The second-order power spectrum
is shown in Fig. 1 for a CDM-like spectrum with shape parameter $\Gamma=0.25$,
with
and the amplitude is chosen such that $\Delta^2(k=0.1) = 0.03$.  
The mass field is smoothed on a scale of 2 $h^{-1}$ Mpc, so that
the linear smoothed variance is $\sigma_0^2=0.20$, and the bias parameters
are $b_1=b_2=b_3=1$.   Note that the nonlinear bias masks the effect seen
in the unbiased case by \scite{JB94} of a transfer of power from 
large to small scales.

The constant power at large scales should not be confused with the 
requirement that $P(k)\rightarrow 0$ as $k\rightarrow 0$, which 
applies if the mean density is estimated from the survey.  This 
will suppress the power for $k \ls 1/L$, where $L$ is the
characteristic depth of the survey, and is quite independent of the
considerations here.  Similarly, the neglect of constant terms in $\delta$
affects only $\bk={\bf 0}$.  

The term here has been conjectured before, and it is appropriate to make a 
few remarks about its nature.  It is essentially a shot-noise term
arising from the peaks {\em and troughs} in the underlying density field
being nonlinearly-biased by the quadratic term $b_2$. The biasing gives
a contribution to the density field consisting of spikes at the peaks and 
troughs. We consider here the case when the underlying matter power spectrum
tends to zero on large scales, so we look only at this remaining 
`spike' contribution (see Appendix A). 
It may seem odd that the troughs make a contribution;  is should be 
remembered that they do not necessarily
correspond to actual overdensities, as there is a linear contribution which
generally lowers the density - a positive quadratic term merely means the 
underdense regions are not quite as underdense as a linear bias would imply.  
However, by assumption, this linear contribution has zero power on very 
large scales, so the quadratic term dominates.
   
To illustrate the shot-noise nature of the spikes, consider a 
non-pathological power spectrum, filtered to remove power with $k>k_{\rm max}$.
It is straightforward to show that the constant-power term is roughly
\be
P^g_{22} \simeq b_2^2 \Delta^4(k_{\rm max}) {1 \over \bar n}
\ee
where $\Delta(k_{\rm max})$ measures the nonlinearity on the smoothing scale,
and $\bar{n} \sim k_{\rm max}^{-3}$ is the number density of peaks and 
troughs on this scale (\pcite{PH85}, \pcite{BBKS}).  Shot noise of a 
point process gives $P_{\rm shot} = 1/\bar n$ (e.g. \pcite{Peebles});  the
prefactor arises here because each spike has a height of roughly
$b_2 \sigma_0^2$ so the integral under each spike is 
$\sim b_2 \Delta^2(k_{\rm max})$.   

This shot noise argument would be dangerous if it was applied to peaks 
alone, as we know that these are strongly clustered 
(\pcite{Kai84}, \pcite{PH85}, 
\pcite{BBKS}), and their clustering might well exceed the shot noise 
contribution. In the appendix we prove that high peaks and low troughs, 
taken together, are very weakly clustered, so the power contribution is indeed
essentially a shot noise term from unclustered small-scale spikes in the 
biased field. 

\begin{figure}
\centering
\begin{picture}(200,200)
\includegraphics{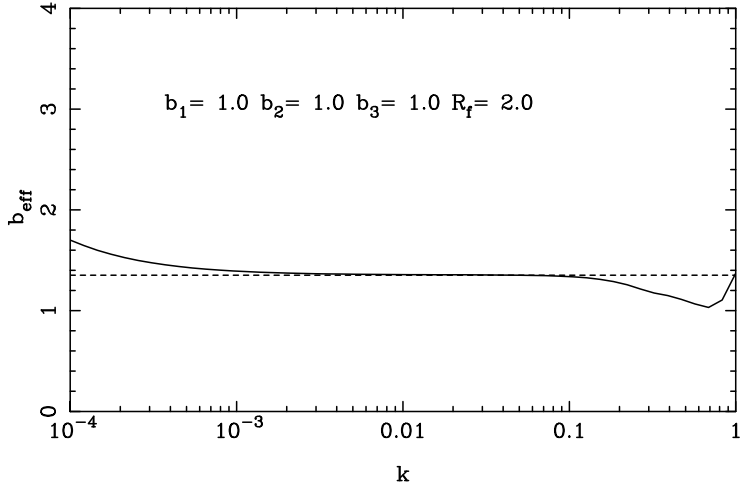}
\end{picture}
\caption[]{\label{RealBias} Effective bias in real space, from perturbation
theory (solid).  This is 
approximately constant over a fairly wide range of $k$, yet the bias is
far from linear.  The dotted line shows the approximate analytic formula 
(\ref{beff}).  The filter erases power at $k\gs 0.2$ and the spike at $k=1$
is due to the matter power spectrum crossing zero, by which time third-order
perturbation theory has broken down.}
\end{figure}

\subsection{Effective bias on large scales is not $b_1$}

If the $P^g_{22}$ term can be neglected in comparison with $P^g_{13}$, then
the effect of biasing and second-order evolution is to add a term which is 
proportional to the linear $P$.   This holds over a wide range of scales, 
but not
at the very largest scales, where the power spectrum is dominated by the
constant contribution from $P^g_{22}$, leading to a divergent bias. 

The fact that  $P^g_{13}$ is proportional to $P_{11}$ gives a constant 
effective bias [$b_{\rm eff}^2 \equiv  P^g(k)/P(k) 
\simeq P^g(k)/P_{11}(k)$].  
In linear theory, the effective bias
is simply $b_1$;  in this section we compute the corrections to this value.
In real space, only 3 terms of $F^{(3)}$ survive.  The first, proportional to
$J^{(3)}$ comes from nonlinear evolution of the mass field, and is very 
small on large scales, and we neglect it here.  There is a straightforward 
contribution from $b_3/6$, which gives 
\be
b_1 b_3 P_{11}(k) \sigma_0^2
\ee
where $\sigma_0^2 = (2\pi^2)^{-1} \int dk k^2 P_{11}(k)$ is the linear
variance in the smoothed field.  The other nonzero term is 
\be
{2 b_2\over (2\pi)^3} \int d^3\bk P_{11}(\bq)\left[J^{(2)}_S(-\bq,\bk)  
+J^{(2)}_S(\bq,\bk)+J^{(2)}_S(\bq,-\bq)\right]
\ee
The last term in brackets is zero because $J$ is zero, 
and the first terms are
equal.  Integration over angles gives a contribution again proportional
to $\sigma_0^2$,  and the final result is
\be
P^g_{13}(k)  =  b_1 P_{11}(k)\left({68\over 21}b_2+b_3\right)\sigma_0^2.
\ee
The $P^g_{13}$ term is of more practical importance than $P^g_{22}$.
We see that it is 
proportional to the linear power spectrum, so it corresponds to a 
scale-independent bias, which, when including the linear terms, gives an
effective bias of
\be
b^2_{\rm eff}(k) = b_1^2 + b_1\left({68\over 21}b_2 + b_3\right)\sigma_0^2.
\label{beff}
\ee
The level of bias depends on the variance of the smoothed field, for which
the truncated Taylor expansion is assumed to be a good description of the
galaxy density field.  It also depends on both the second- and the 
third-order bias coefficients.  The suggestion that the bias should be constant
on large scales has been made by a number of authors 
(\pcite{Coles93}; \pcite{SW97}; \pcite{MPH98}), and we confirm that 
this is true to third order in perturbation theory, provided one is not looking
at the very largest scales where $P^g_{22}$ dominates.

The effective bias 
is plotted in Fig. 2,  along with the prediction of (\ref{beff}), which is 
seen to be a good approximation over a wide range of $k$, for these 
parameters.  
In Fig. 3 we demonstrate that the approximate formula works well for a biased 
N-body simulation, with the same
linear power spectrum, normalised to $\sigma_8=0.26$, 
from the Hydra consortium \cite{Hydra}.  The rather small amplitude of
the power spectrum gives us a large dynamic range between the smoothing scale
and the size of the simulation.  The smoothing
is done with a Gaussian filter of radius $1\ h^{-1}$ Mpc, giving a variance 
of $\sigma_0^2=0.17$, and we take $b_1=b_2=b_3=1$.  Note that a bias 
function which includes non-zero $b_i$ terms beyond $i=3$, such as
the exponential high-peak biasing $\exp(\alpha \delta)$ (\pcite{PW84},
\pcite{JS86}, \pcite{BBKS}) gives
essentially the same result provided that the variance in the field is not
too large.  For the power spectra here, this requires $\alpha \sigma_0 \ls 
0.3$,  but the exact figure depends on how skewed the field is.
As the field becomes
progressively more nonlinear, the bias increases slightly towards high $k$,
and the approximate formula (\ref{beff}) begins to underestimate the bias.
Note that the simulations do not probe a large enough scale to detect the
constant contribution to the power.

Since $P^g_{13}$ is proportional to the linear power, it is formally possible
at this level of perturbation theory to cancel the linear power altogether 
by appropriate choice of $b_2$ and $b_3$, leaving the constant $P^g_{22}$ term.
Perhaps not surprisingly,  it appears impossible to do this on N-body 
simulations without perturbation theory breaking down.   
\begin{figure}
\centering
\begin{picture}(200,200)
\includegraphics{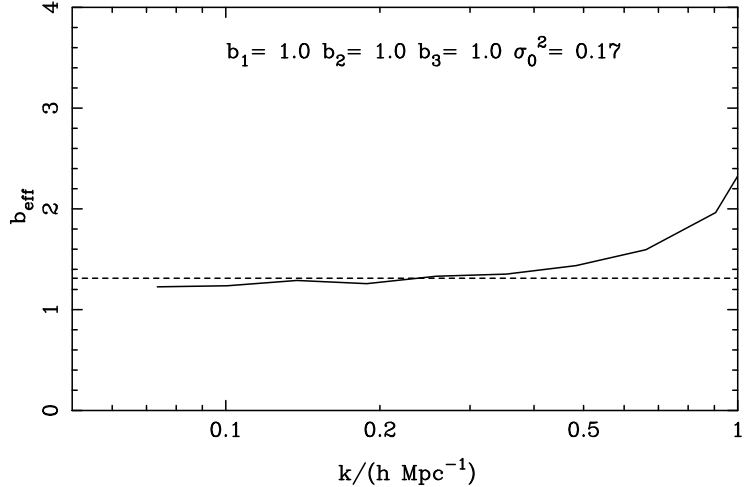}
\end{picture}
\caption[]{\label{RealHydra}  The effective bias factor $\sqrt{
P_{biased}/P_{mass}}$ for a Hydra N-body simulation (solid), along with the
approximate formula (\ref{beff}), for $b_1=b_2=b_3=1$ and $R_f=2 h^{-1}$ Mpc,
which gives an r.m.s. fractional overdensity of 0.41.  For other details, 
see text. This figure also differs from Fig. 2 in that the biased field 
has been deconvolved with a Gaussian, which affects the results at high $k$.}

\end{figure}

\section{Redshift-space}

In Fig. 4 we show the effects in redshift space of biasing and evolution,
for different angles of wavevectors with the line-of-sight, showing similar
qualitative behaviour to Kaiser's (1984) analysis.  
In Fig. 5 we show the redshift-space power spectrum (averaged over angle) 
for the Hydra simulation, for the same smoothings and parameters as in Fig. 3.
 We see here remarkably good agreement with perturbation theory of biased 
fields in redshift space.  Note that linear theory, with linear bias, predicts
that the redshift-space and real-space power spectra should be proportional
to each other.  Here we see that the higher-order terms give very different
behaviour.  Note also that perturbation theory gives excellent results all 
the way down to Mpc scales, where the smoothing is applied.
\begin{figure}
\centering
\begin{picture}(200,200)
\includegraphics{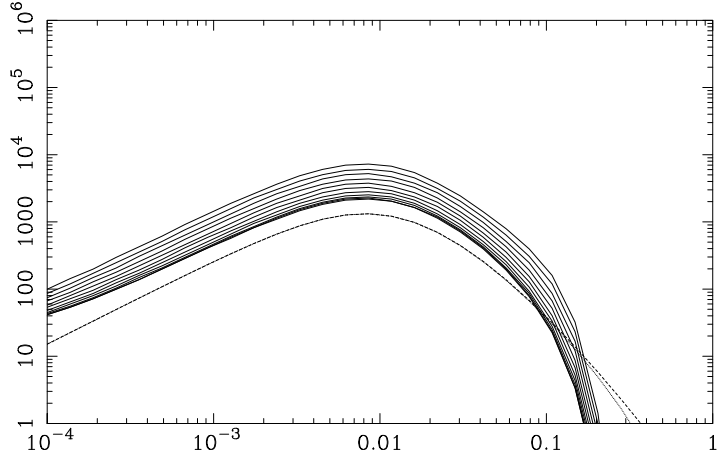}
\end{picture}
\caption[]{\label{RedPower} Redshift-space power spectrum for the model 
of Fig. 1.  The curves are for values of $\mu$ from 0 (bottom) to 1 in 
steps of 0.1.
The dashed and dotted lines are the unsmoothed linear and nonlinear power
spectrum in real space.}
\end{figure}
\begin{figure}
\centering
\begin{picture}(200,200)
\includegraphics{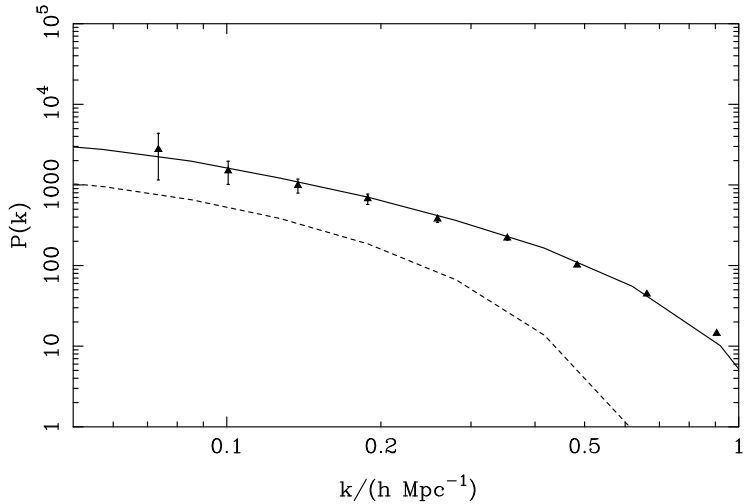}
\end{picture}
\caption[]{\label{RedHydra} Hydra simulations in redshift space (points), 
along with the theoretical curve from perturbation theory 
($b_1=b_2=b_3=1.0$, and $\sigma_0=0.41$).  The dotted line
is the smoothed linear theory matter power spectrum in real space.}
\end{figure}
\subsection{Quadrupole-to-monopole ratio}

\begin{figure}
\centering
\begin{picture}(200,200)
\includegraphics{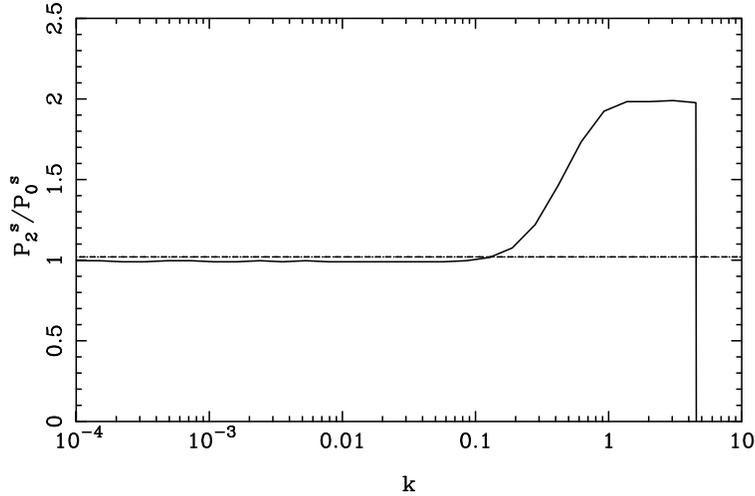}
\end{picture}
\caption[]{\label{QuadToMonMass} Quadrupole-to-monopole ratio for the mass, 
as a function of wavenumber (solid).  Beyond $k=4.5$, the filtered power is
zero to machine accuracy.  The dashed line is the linear theory ratio 
(\ref{QtoM}).}
\end{figure}

\begin{figure}
\centering
\begin{picture}(200,200)
\includegraphics{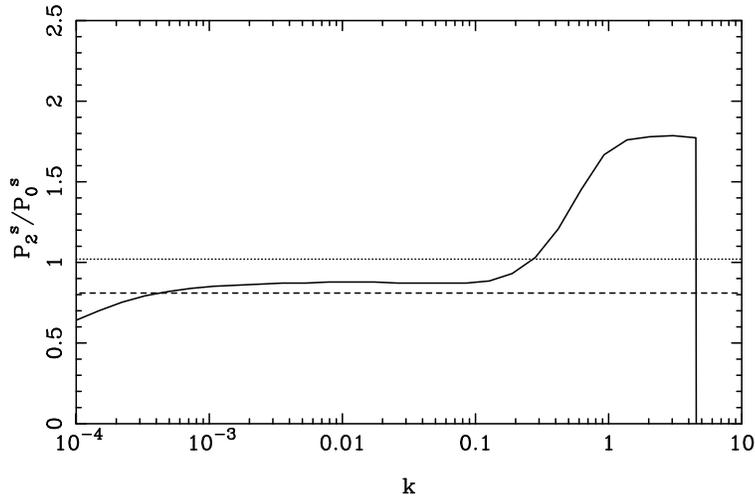}
\end{picture}
\caption[]{\label{QuadToMon} Quadrupole-to-monopole ratio of the biased 
field (solid), as a function of
wavenumber, along with unbiased linear theory (dotted) and linear theory,
but with $b$ replaced by $b_{\rm eff}$ from perturbation theory (dashed).
The behaviour at $k<10^{-3}$ arises from the constant-power term in the biased
field.}
\end{figure}

\scite{Ham92} proposed the quadrupole-to-monopole 
ratio of the power spectrum as a diagnostic for $\beta$.  These are obtained 
by expansion of the power spectrum in a shell of fixed $|\bk|$, in terms of 
Legendre polynomials (in $\mu$) of order 0 and 2.    In linear theory,
the ratio of the coefficients ${\cal P}_i^s$  has the value
\be
{{\cal P}_2^s\over {\cal P}_0^s} = 
{ {4\over 3}\beta + {4\over 7}\beta^2 \over 1+{2\over 3}
\beta + {1\over 5}\beta^2} \;.
\label{QtoM}
\ee
\scite{TayHam96} extended this idea by using the
Zel'dovich approximation to investigate the behaviour of the 
quadrupole-to-monopole ratio into the mildly nonlinear regime. In this 
section we use perturbation theory results to investigate the influence
that bias might have on conclusions drawn from the ratio.  In practice,
it is best to use the full 3D power spectrum to estimate parameters, but
analysis of the quadrupole-to-monopole ratio can be a useful aid to see
why we might be able to extract information, and why there might be 
problems.  In Fig. \ref{QuadToMonMass} we see that, on large scales, 
linear theory accounts well for the ratio, with significant deviations 
appearing on small scales in qualitative agreement with \scite{TayHam96}.  
The slight discrepancy with linear theory may arise from only 
computing the power spectrum with
a rather crude separation of $\Delta\mu=0.1$.
On small scales, the quadrupole-to-monopole ratio was studied by 
\scite{TayHam96} and
\scite{HC98}, where it is affected seriously by
caustic formation and small-scale virialised structures.  
We confine our remarks to larger scales ($k \ls 0.1$),  where these effects
are not important, but there are still some surprises.
Fig. \ref{QuadToMon} shows the quadrupole-to-monopole ratio for
a biased CDM-like field, with $\Gamma=0.25$, with $b_1=b_2=b_3=1.0$ and 
$\sigma_0^2=0.196$ and $R_f=2\,h^{-1}$ Mpc.  There is a
large range of wavenumbers for which the ratio is very nearly constant,
which might give one confidence that a linear bias is an adequate 
description.  However, the retrieved value of $\beta=0.81$ 
agrees neither with linear theory ($\beta=1$), nor with the appropriate
$\beta$ from the effective bias on large scales ($\beta=0.74$).  It is the
latter which we would like to estimate accurately, since it is the effective 
bias which tells us the amplitude of the underlying matter power spectrum.
In this case, the recovered effective bias (1.24) is not close to the
true effective bias (1.35).  The discrepancy arises because
the one-loop corrections do not have the same angular dependence
as the Kaiser factor multiplying the linear power.  This is not really a
breakdown of linear theory as such;  the underlying mass field is still
well described by linear perturbation theory.  Here it is the nonlinear
bias which causes the problem, and the simple analysis of \scite{Kai87}
cannot be applied.  These findings mean
that attempts to measure $\beta$ from linear redshift distortions (e.g.
\pcite{Ham92}; \pcite{FSL94}; \pcite{HT95}; \pcite{BHT95}; \pcite{CFW95};
\pcite{Tad98}) 
must make the further assumption that the one-loop corrections are small.

\section{Conclusions}

In this paper, we have shown how the power spectrum is altered in real and 
redshift space by the effects of nonlinear local bias and 
evolution.  The assumptions in the analysis are that the smoothed galaxy
density field is some local function of the underlying smoothed present-day
mass field (i.e. Eulerian local bias).  
The smoothing scale is assumed to be large enough that a Taylor expansion 
to third order:
$\delta_g \simeq \sum_{j=0}^3 b_j \delta^j/j!$ suffices.  
The approach we have used is perturbative; the one-loop 
corrections give contributions from third-order (in $\delta$) which are as 
important as second order, so we have kept evolution and bias terms to 
this order ($j=3$).
We confirm the predictions of \scite{SW97} that the power spectrum has a 
constant component, plus a roughly 
scale-independent bias on large scales.  We show
that the constant power term is essentially a shot noise term arising from
quadratic biasing of peaks and troughs.  The
most important results in real space are the expressions for the magnitudes 
of the two effects: equation (\ref{beff}) for the effective 
bias on large scales,
and (\ref{DC}) for the constant contribution, which could in principle
make difficult the task of unveiling the underlying 
power spectrum on extremely large scales. However, our 
estimates of the size of this effect suggest that it is not going to be a 
practical problem in the foreseeable future,  since it is likely to dominate 
only on scales of the order of the horizon size.  Should the galaxy 
power spectrum be measured on extremely large scales at some time in the 
future,  there is a possibility that a deviation from the scale-invariant
spectrum (in $|k|$) might be interpreted as evidence for an open Universe
\cite{KS94}.  This paper provides an alternative interpretation in terms
of nonlinear bias.

Potentially more serious is the effect of nonlinear bias on the amplitude of
the power spectrum, in both real and redshift space.  Here there is the
possibility of confusion, in that the shape of the linear power spectrum is
preserved over a wide range of $k$, so one can define a scale-independent 
effective bias from the ratio of the galaxy to matter power spectra.
The difficulty is that this bias is not the linear bias ($b_1$ in the 
Taylor expansion of the galaxy density field), nor is it the bias which
would be recovered from studying anisotropy in the redshift power spectrum.
These complications may compromise efforts to deduce cosmological parameters
from redshift distortions or peculiar velocities, unless a further assumption
is made that the one-loop corrections are small.  Without a better 
understanding of galaxy formation, it is impossible to make this assumption
with a great deal of confidence.  What would be interesting to see is
whether current models can identify a scale on which an Eulerian
bias is a good description of the galaxy field;  if such a scale can be
identified, then the computations in this paper may be useful in providing 
analytic approximations for the effects on the power spectrum and 
redshift distortion. 

\section* {Acknowledgments}

LV acknowledges support from the European Union for support under the
TMR programme.  AFH thanks the University 
of Padova, and SM the University of Edinburgh for hospitality.
Calculations were performed on STARLINK facilities.  The simulations 
were obtained from the data bank of cosmological N-body simulations 
provided by the Hydra consortium (http://coho.astro.uwo.ca/pub/data.html) 
and produced using the Hydra N-body code (Couchman et al. 1995). We thank 
Francesco Lucchin, Lauro Moscardini, Bhuvnesh Jain and Andy Taylor for 
useful discussions.

\bibliographystyle{mnras}
\bibliography{general}

\appendix
\section{Correlations of high and low regions}

A series expansion for the correlation function of regions above a threshold
was obtained by \scite{JS86}; see also \scite{PW84}.  
The first term in this series is the one
obtained by \scite{Kai84}, which shows that high regions can be very strongly
clustered.  In this appendix, we calculate the correlation
function of high regions (above a threshold) plus troughs 
(below a threshold), and
show that high regions plus troughs are very weakly correlated on large scales,
with a correlation function which is proportional to the square of the 
correlation function of the field.

We follow the formalism and notation of \scite{JS86}.  Let the variance in
the Gaussian field be $\sigma^2$, and consider high regions above 
$\delta = \nu \sigma$ and troughs below $\delta =-\nu \sigma$.  

The two-point correlation function $\xi_X$ of regions separated by $r$,
with $\nu$ in some
region $X$ is given by an integral over the normal bivariate Gaussian:
\be
1+\xi_X = {I^{-2}\over 2 \pi {\rm det}C } \int_X \int_X d \nu_1 d \nu_2 
\exp\left[-{1\over 2}
\nu_i C_{ij}^{-1}\nu_j\right] 
\ee
where
\be
I \equiv \int_X {d \nu \over \sqrt{2\pi}} \exp\left(-\nu^2/2\right)
\ee
and the covariance matrix $C_{ij} \equiv \langle \nu_i \nu_j \rangle$
is unity on the diagonal, and $\psi = \xi(r)/\sigma^2$ off the diagonal.

To evaluate the double integral, we use the same trick as \scite{JS86}.  We
take a Fourier transform of the exponential:
\be
\exp\left[-{1\over 2}\left(
\nu_i C_{ij}^{-1}\nu_j \right)\right] = \int {d^2k\over 2 \pi} \exp(-i {\bf k}
\cdot {\bf \nu}) \exp \left[-{1\over 2}\left(k_i C_{ij} k_j\right)\right].
\ee
Then the double integral may be written
\be
\int_X \int_X d\nu_1 d \nu_2 \int {d^2k\over 2 \pi} \exp(-i {\bf k}
\cdot {\bf \nu}) \exp(-k_i k_i/2)\exp(Q),
\ee
where $Q = -k_1 k_2 \psi$.  We expand $\exp(Q)=\sum_{m=0}^\infty Q^m/m!$, 
and then regard $Q$ as an operator 
\be
Q\exp(-i {\bf k}\cdot {\bf \nu}) = \psi {\partial\over \partial \nu_1} 
{\partial\over \partial \nu_2}\exp(-i {\bf k}\cdot {\bf \nu}).
\ee
The double integral can then be written
\be
\int_X \int_X d\nu_1 d \nu_2 \sum_{m=0}^\infty {\psi^m\over m!}
\left( {\partial\over \partial \nu_1} 
{\partial\over \partial \nu_2} \right)^m  
\exp\left[-(\nu_1^2 + \nu_2^2)/2\right].
\ee
\scite{JS86} took the region to be $\nu_i > \nu$.  We take the region
to be  $|\nu_i| > \nu$, which gives four contributions to the double integral,
and $I$ is also modified to twice the high region result, i.e. 
${\rm erfc}(\nu/\sqrt{2})$.  The terms in the double integral from peak-peak
and trough-trough give the same as \scite{JS86}, i.e.
\be
2 \sum_{m=0}^\infty  {\psi^m\over m!} \left[H_{m-1}(\nu) 
\exp(-\nu^2/2)\right]^2,
\ee
where the Hermite polynomial is 
$H_m(x) \equiv \exp(x^2/2) (-d/dx)^m \exp(-x^2/2)$.
The cross-terms in the double integral give an identical contribution,
but multiplied by $(-1)^m$.  The effect of this is to eliminate the
odd terms in the series, whilst leaving the even terms intact.  The 
strong clustering term (for high regions alone) which Kaiser 
computed disappears, leaving the first non-zero 
contribution as
\be
\xi_X = {\nu^4\over 2} \psi^2 + O(\psi^4)
\ee
so if the underlying density field has small correlations on large scales
($\psi \rightarrow 0$), the field consisting of high regions and low 
troughs is very weakly clustered.

\end{document}